\documentclass[12pt]{article}
\usepackage{amsfonts}
\usepackage{amssymb}
\usepackage{enumerate}
\usepackage{amsmath}

\newcounter{defin}  \newcounter{lemma}  \newcounter{theorem}
\newcounter{property} \newcounter{corol}  \newcounter{remark}
\newcounter{example}

\newenvironment{theorem}{\par\refstepcounter{theorem}%\noindent
     \textbf{Theorem \thetheorem.}\ }{\rm\par}
\newenvironment{property}{\par\refstepcounter{property}%\noindent
     \textbf{Proposition \theproperty.}\ }{\rm\par}
\newenvironment{corollary}{\par\refstepcounter{corol}%\noindent
     \textbf{Corollary \thecorol.} }{\rm\par}
\newenvironment{definition}{\par\refstepcounter{defin}%\noindent
     \textbf{Definition \thedefin.}\ }{\rm\par}
\newenvironment{remark}{\par\refstepcounter{remark}%\noindent
     \textbf{Remark \theremark.}}{\rm\par}
\newenvironment{example}{\par\refstepcounter{example}%\noindent
     \textbf{Example \theexample.}}{\rm\par}

\begin{document}
\title{On channels with finite Holevo capacity.}
\author{M.E.Shirokov \thanks{Steklov Mathematical Institute, 119991 Moscow,
Russia}}
\date{}
\maketitle
\section{Introduction}

As it is well known, the properties of quantum channels differ
quite substantially for infinite and finite quantum systems. We
consider a nontrivial class of infinite dimensional quantum
channels, characterized by finiteness of the Holevo capacity (in
what follows the $\chi$-capacity).

The results in  \cite{Sh-4} imply some general properties of
channels with finite $\chi$-capacity such as compactness of the
output set and existence of the unique output optimal average
state (theorem \ref{o-a-s}). It is shown that each channel with
finite $\chi$-capacity can be uniformly approximated by channels
with finite dimensional output system, having close
$\chi$-capacities and the output optimal average states (corollary
\ref{approximation}).

The class of channels with finite $\chi$-capacity contains
subclass of channels with continuous output entropy, called
$\textup{CE}$-channels. These channels inherit some analytical
properties of finite dimensional channels (proposition
\ref{chi-function}), revealing at the same time essential features
of infinite dimensional channels (example \ref{example-2}). So, we
may consider the class of $\textup{CE}$-channels as an
intermediate one and use technical advantages of dealing with such
channels to approach general infinite dimensional channels and
systems. For example, the channel of this type was used in
\cite{H-Sh-W} in considering properties of infinite dimensional
entanglement-breaking channels and of the set of separable states.

In \cite{H-Sh-2} the notion of an optimal measure (generalized
ensemble) for constrained infinite dimensional channels was
introduced and the sufficient condition of its existence was
obtained. In this paper we consider a special condition of
existence of an optimal measure for unconstrained infinite
dimensional channels with finite $\chi$-capacity (theorem
\ref{o-m}, corollary \ref{o-m-s-c}, remark \ref{on o-m-s-c}) and
construct examples of channel with bounded output entropy, for
which there exist no optimal measures (examples \ref{example-1}C
and \ref{example-2}).

An interesting question is a problem of extension of a quantum
channel from the set of all quantum states (=normal states) on
$\mathfrak{B}(\mathcal{H})$ to the set of all states (=positive
normalized linear functionals) on $\mathfrak{B}(\mathcal{H})$,
which is a compact set in the $*$-weak topology. By using a simple
topological observation we obtain a condition of existence of an
extension of a channel to nonnormal states (proposition
\ref{compact output}), which implies existence of such extension
for arbitrary channel with finite $\chi$-capacity (corollary
\ref{c-f-c-ext}). The entropic characteristics of a channel
extension are considered as well as the condition of existence of
an optimal measure (proposition \ref{o-m-ext}), which shows the
meaning of the conditions in theorem \ref{o-m} (remark \ref{on
i-o-a-s}).

We complete this paper by considering the class of infinite
dimensional channel, for which the $\chi$-capacity can be
explicitly determined and there exists a simple condition of
continuity of the output entropy (proposition
\ref{orbit-channel}). The $\textup{CE}$-channels of this class
demonstrate essential infinite dimensional features such as
nonexistence of an input optimal average state and of an optimal
measure, discontinuity of the $\chi$-capacity as a function of a
channel (example \ref{example-2}).

\section{Preliminaries}

Let $\mathcal{H}$ be a separable Hilbert space,
$\mathfrak{B}(\mathcal{H})$ - the Banach space of all bounded
operators in $\mathcal{H}$ with the cone
$\mathfrak{B}_{+}(\mathcal{H})$ of all positive operators,
$\mathfrak{T}( \mathcal{H})$ - the Banach space of all trace-class
operators with the trace norm $\Vert\cdot\Vert_{1}$ and
$\mathfrak{S}(\mathcal{H})$ - the closed convex subset of
$\mathfrak{T}(\mathcal{H})$ consisting of all density operators in
$\mathcal{H}$, which is complete separable metric space with the
metric defined by the trace norm. Each density operator uniquely
defines a normal state on $\mathfrak{B}(\mathcal{H})$ \cite{B&R}.
We will also consider the set
$\widehat{\mathfrak{S}}(\mathcal{H})$ of all normalized positive
functionals on $\mathfrak{B}(\mathcal{H})$, which is a compact
subset of $(\mathfrak{B}(\mathcal{H}))^{*}$ in the $*$-weak
topology.  The set $\mathfrak{S}(\mathcal{H})$ can be considered
as a $*$-weak dense subset of
$\widehat{\mathfrak{S}}(\mathcal{H})$ \cite{B&R}. Since the main
attention in this paper is focused on $\mathfrak{S}(\mathcal{H})$
rather then $\widehat{\mathfrak{S}}(\mathcal{H})$ we will use the
term \textit{state} for elements of $\mathfrak{S}(\mathcal{H})$
while the elements of
$\widehat{\mathfrak{S}}(\mathcal{H})\setminus\mathfrak{S}(\mathcal{H})$
will be called \textit{nonnormal states}.

Let $\mathcal{H},\mathcal{H}^{\prime }$ be a pair of separable
Hilbert spaces which we shall call correspondingly input and
output space. A channel $\Phi $ is a linear positive trace
preserving map from $\mathfrak{T}(\mathcal{ H })$ to
$\mathfrak{T}(\mathcal{H}^{\prime })$ such that the dual map $\Phi
^{\ast }:\mathfrak{B}(\mathcal{H}^{\prime
})\mapsto\mathfrak{B}(\mathcal{H})$ (which exists since $\Phi $ is
bounded \cite{H-SSQT}) is completely positive.

For arbitrary set $\mathcal{A}$ let $\mathrm{co}(\mathcal{A})$ and
$\overline{\mathrm{co}}(\mathcal{A})$ be the convex hull and the
convex closure of the set $\mathcal{A}$ correspondingly and let
$\mathrm{Ext}(\mathcal{A})$ be the set of all extreme points of
the set $\mathcal{A}$ \cite{J&T}.

We will use the following compactness criterion for subsets of
states: \textit{a closed subset $\mathcal{K}$ of states is compact
if and only if for any $\varepsilon>0$ there is a finite
dimensional projector $P$ such that $\mathrm{Tr}\rho P\geq
1-\varepsilon$ for all $\rho\in\mathcal{K}$}
\cite{S},\cite{H-Sh-2}.

Speaking about continuity of a particular function on some set of
states we mean continuity of the restriction of this function to
this set.

Arbitrary finite collection $\{\rho _{i}\}$ of states in
$\mathfrak{S}(\mathcal{H})$ with corresponding set of
probabilities $\{\pi _{i}\}$ is called \textit{ensemble} and is
denoted by $\{\pi _{i},\rho _{i}\}$. The state
$\bar{\rho}=\sum_{i}\pi _{i}\rho _{i}$ is called \textit{the
average state} of the ensemble. Following \cite{H-Sh-2} we treat
an arbitrary Borel probability measure $\mu$ on
$\mathfrak{S}(\mathcal{H})$ as \textit{generalized ensemble} and
the barycenter  of the measure $\mu$ defined by the Pettis
integral
\[
\bar{\rho}(\mu )=\int\limits_{\mathfrak{S}(\mathcal{H})}\rho \mu
(d\rho )
\]
as \textit{the average state} of this ensemble. In this notations
the conventional ensembles correspond to measures with finite
support. For arbitrary closed subset $\mathcal{A}$ of
$\mathfrak{S}(\mathcal{H})$ we denote by
$\mathcal{P}(\mathcal{A})$ the set of all probability measures
supported by the set $\mathcal{A}$ \cite{Par}. In what follows an
arbitrary ensemble $\{\pi _{i},\rho _{i}\}$ is considered as a
particular case of probability measure and is also denoted by
$\mu$.

Consider the functionals
$$
\chi_{\Phi}(\mu)=\int
H(\Phi(\rho)\Vert\Phi(\bar{\rho}(\mu)))\mu(d\rho)\quad\mathrm{and}\quad
\hat{H}_{\Phi}(\mu)=\int H(\Phi(\rho))\mu(d\rho).
$$
In \cite{H-Sh-2} (proposition 1 and the proof of the theorem) it
is shown that both these functionals are lower semicontinuous on
$\mathcal{P}(\mathfrak{S}(\mathcal{H}))=\mathcal{P}$ and
\begin{equation}\label{formula}
\chi_{\Phi}(\mu)=H(\Phi(\bar{\rho}(\mu)))-\hat{H}_{\Phi}(\mu)
\end{equation}
for arbitrary $\mu$ such that $H(\Phi(\bar{\rho}(\mu)))<+\infty$.

If $\mu=\{\pi_{i},\rho_{i}\}$ then
$$
\chi_{\Phi}(\{\pi_{i},\rho_{i}\})=\sum_{i=1}^{n}\pi_{i}H(\Phi(\rho_{i})\Vert
\Phi(\bar{\rho}))\quad\mathrm{and}\quad
\hat{H}_{\Phi}(\{\pi_{i},\rho_{i}\})=\sum_{i=1}^{n}\pi_{i}H(\Phi(\rho_{i})).
$$

In what follows we will use the \textit{decrease coefficient}
$\mathrm{dc}(\sigma)$ of a state $\sigma$ defined in \cite{Sh-4}
by
$$
\mathrm{dc}(\sigma)=\inf\{\lambda>0\,|\,\mathrm{Tr}\sigma^{\lambda}<+\infty\}\in[0,1].
$$

We will use the notion of the \textit{$H$-convergence} of a
sequence of states $\{\rho_{n}\}$ to a state $\rho_{0}$ defined by
the condition $\lim_{n\rightarrow+\infty}H(\rho_{n}\|\rho_{0})=0$.

In \cite{Sh-4} the properties of the $\chi$-capacity as a function
of a set of states were explored and the notion of the optimal
average state $\Omega(\mathcal{A})$ of a set $\mathcal{A}$ with
finite $\chi$-capacity was introduced. The set of states
$\mathcal{A}$ with finite $\chi$-capacity is called regular if one
of the two following conditions hold:
\begin{itemize}
\item $H(\Omega(\mathcal{A}))$ is finite and $\lim_{n\rightarrow+\infty}H(\rho_{n})=H(\Omega(\mathcal{A}))$
       for arbitrary sequence $\{\rho_{n}\}$ of states in
       $\mathrm{co}(\mathcal{A})$ $H$-converging to the state
       $\Omega(\mathcal{A})$;
  \item the function $\rho\mapsto H(\rho\|\Omega(\mathcal{A}))$
                is  continuous on the set $\overline{\mathcal{A}}$.
\end{itemize}

In a sense these conditions are the minimal continuity
requirements which guarantee the "good" properties of the
$\chi$-capacity (see theorems 2 and 3 in \cite{Sh-4}). The
simplest sufficient condition of regularity of a set $\mathcal{A}$
is given by the inequality $\mathrm{dc}(\Omega(\mathcal{A}))<1$
(theorem 2E in \cite{Sh-4}), but this condition is not necessary
and there exist regular sets, consisting of states with infinite
entropy.

The all notations used in this paper coincide with the notations
accepted in \cite{Sh-4}.

\section{General properties}

Let
$\Phi:\mathfrak{S}(\mathcal{H})\mapsto\mathfrak{S}(\mathcal{H}')$
be a channel with finite $\chi$-capacity defined by
\begin{equation}\label{ccap-1}
\bar{C}(\Phi)=\sup_{\{\pi_{i},\rho_{i}\}}\chi_{\Phi}(\{\pi_{i},\rho_{i}\})
=\sup_{\{\pi_{i},\rho_{i}\}}\sum_{i}\pi_{i}
H(\Phi(\rho_{i})\|\Phi(\bar{\rho})).
\end{equation}

In \cite{H-Sh-2} it is shown that
\begin{equation}\label{ccap-2}
\bar{C}(\Phi)=\sup_{\mu\in\mathcal{P}}\chi_{\Phi}(\mu)=
\sup_{\mu\in\mathcal{P}}
\int\limits_{\mathfrak{S}(\mathcal{H})}H(\Phi(\rho)\Vert\Phi(\overline{\rho}(\mu))\mu(d\rho).
\end{equation}

According to \cite{Sh-2} a sequence of ensembles
$\{\{\pi_{i}^{n},\rho _{i}^{n}\}\}_{n}$ such that
$$
\lim_{n\rightarrow+\infty}\sum_{i}\pi^{n}_{i}H(\Phi(\rho^{n}_{i})\|\Phi(\bar{\rho}_{n}))=\bar{C}(\Phi)
$$
is called \textit{approximating sequence} for the channel $\Phi$
while any partial limit of the corresponding sequence of the
average states
$\{\bar{\rho}_{n}=\sum_{i}\pi_{i}^{n}\rho_{i}^{n}\}$ is called
\textit{input optimal average state} for the channel $\Phi$.

By the compactness argument for a finite dimensional channel there
exists at least one input optimal average state, coinciding with
the average state of the optimal ensemble for this channel
\cite{Sch-West-1}. For an infinite dimensional channel with finite
$\chi$-capacity existence of an input optimal average state is a
question depending on a channel (example \ref{example-1}), which
is closely related to the question of existence of an optimal
measure for this channel (theorem \ref{o-m}, corollary
\ref{o-m-s-c}).

By the definitions the $\chi$-capacity of a channel $\Phi$
coincides with the $\chi$-capacity of the output set
$\Phi(\mathfrak{S}(\mathcal{H}))$ of this channel. Thus theorems 1
and 2D in \cite{Sh-4} imply the following observation.

\begin{theorem}\label{o-a-s}
\textit{Let
$\Phi:\mathfrak{S}(\mathcal{H})\mapsto\mathfrak{S}(\mathcal{H}')$
be a channel with finite $\chi$-capacity.}

\textit{The set $\Phi(\mathfrak{S}(\mathcal{H}))$ is a relatively
compact subset of $\mathfrak{S}(\mathcal{H}')$.}

\textit{There exists the unique state $\Omega(\Phi)$ in
$\mathfrak{S}(\mathcal{H}')$ such that
$$
H(\Phi(\rho)\|\Omega(\Phi))\leq \bar{C}(\Phi), \quad \forall
\rho\in \mathfrak{S}(\mathcal{H}).
$$}\vspace{-15pt}

\textit{The state $\Omega(\Phi)$ lies in
$\overline{\Phi(\mathfrak{S}(\mathcal{H}))}$. For arbitrary
approximating sequence of ensembles
$\{\{\pi_{i}^{n},\rho_{i}^{n}\}\}_{n}$ the corresponding sequence
$\{\Phi(\bar{\rho}_{n})\}_{n}$ of images of their average states
$H$-converges to the state $\Omega(\Phi)$.}

\textit{If there exists an input optimal average state $\rho_{*}$
for the channel $\Phi$ then $\Phi(\rho_{*})=\Omega(\Phi)$. }

\textit{The $\chi$-capacity of the channel $\Phi$ can be defined
by the expression\footnote{By corollary 9 in \cite{Sh-4} the
infinum in this expression can be over the subset of
$\overline{\Phi(\mathfrak{S}(\mathcal{H}))}$ consisting of states
invariant for all automorphism $\alpha$ of
$\mathfrak{S}(\mathcal{H}')$ such that
$\alpha(\overline{\Phi(\mathfrak{S}(\mathcal{H}))})\subseteq\overline{\Phi(\mathfrak{S}(\mathcal{H}))}$.}
$$
\bar{C}(\Phi)=\inf_{\sigma\in\mathfrak{S}(\mathcal{H}')}\sup_{\rho\in
\mathfrak{S}(\mathcal{H})}H(\Phi(\rho)\|\sigma)=\sup_{\rho\in
\mathfrak{S}(\mathcal{H})}H(\Phi(\rho)\|\Omega(\Phi)).
$$}
\end{theorem}

According to \cite{Sh-2} the state $\Omega(\Phi)$ is called
\textit{the output optimal average state} for the channel $\Phi$.

Below we consider examples of channels with finite $\chi$-capacity
with no input optimal average state, for which the output optimal
average states are explicitly determined and play important role
in studying of these channels.

For a finite dimensional channel $\Phi$ the state $\Omega(\Phi)$ is
an image of the average state of any optimal ensemble for this
channel, which implies
$$
\Omega(\Phi)\in\Phi(\mathfrak{S}(\mathcal{H}))\quad
\mathrm{and}\quad \bar{C}(\Phi)\leq H(\Omega(\Phi)).
$$
For an infinite dimensional channel $\Phi$ with finite
$\chi$-capacity these relations do not hold in general (see example
\ref{example-1}C below). The first relation follows from existence
of at least one input optimal average state for the  channel $\Phi$
while a sufficient condition of the second one can be expressed in
terms of the spectrum of the state $\Omega(\Phi)$ (see proposition
\ref{sr-o-a-s} below).

The simplest examples of infinite dimensional channels with finite
$\chi$-capacity are channels from an infinite dimensional quantum
system into a finite dimensional one. Following \cite{Sh-2} such
channels will be called $\textup{IF}$-channels. Theorem
\ref{o-a-s} implies, in particular, that each channel with finite
$\chi$-capacity can be uniformly approximated by
$\textup{IF}$-channels.
\begin{corollary}\label{approximation}
\textit{The channel
$\Phi:\mathfrak{S}(\mathcal{H})\mapsto\mathfrak{S}(\mathcal{H}')$
has finite $\chi$-capacity if and only if there exists a sequence
$\{\Phi_{n}:\mathfrak{S}(\mathcal{H})\mapsto\mathfrak{S}(\mathcal{H}'_{n}),
\mathcal{H}'_{n}\subseteq\mathcal{H}'\}$ of $\textup{IF}$-channels
such that
$$
\lim_{n\rightarrow+\infty}\sup_{\rho\in\mathfrak{S}(\mathcal{H})}\|\Phi_{n}(\rho)-\Phi(\rho)\|_{1}=0\quad
and \quad\sup_{n}\bar{C}(\Phi_{n})<+\infty.
$$}

\textit{The sequence $\{\Phi_{n}\}$ can be chosen in such a way
that
$$
\lim_{n\rightarrow+\infty}\bar{C}(\Phi_{n})=\bar{C}(\Phi)\quad and
\quad \lim_{n\rightarrow+\infty}\Omega(\Phi_{n})=\Omega(\Phi).
$$
}\end{corollary}

\textbf{Proof.} If the above sequence $\{\Phi_{n}\}$ exists then the
$\chi$-capacity of the channel $\Phi$ is finite due to lower
semicontinuity of the $\chi$-capacity as a function of a channel
\cite{Sh-2}.

If the $\chi$-capacity of the channel $\Phi$ is finite then by
theorem \ref{o-a-s} the set $\Phi(\mathfrak{S}(\mathcal{H}))$ is
relatively compact. Let $\{P_{n}\}$ be a sequence of finite rank
projectors in $\mathcal{H}'$ strongly converging to
$I_{\mathcal{H}'}$. The compactness criterion implies
$\lim_{n\rightarrow+\infty}\inf_{\rho\in\mathfrak{S}(\mathcal{H})}\mathrm{Tr}\Phi(\rho)P_{n}=1$.
Hence the sequence of $\textup{IF}$-channels
$$
 \Phi_{n}(\rho)=P_{n}\Phi(\rho)P_{n}+
\left(\mathrm{Tr}(I_{\mathcal{H}'}-P_{n})\Phi(\rho)\right)\tau_{n}
$$
from $\mathfrak{S}(\mathcal{H})$ into
$\mathfrak{S}(P_{n}(\mathcal{H}')\oplus\mathcal{H}''_{n})$, where
$\tau_{n}$ is a pure state in some finite dimensional subspace
$\mathcal{H}''_{n}$ of $\mathcal{H}'\ominus P_{n}(\mathcal{H}')$,
uniformly converges to the channel $\Phi$. Since for each $n$ the
channel $\Phi_{n}$ can be represented as a composition of the
channel $\Phi$ and the channel $\rho\mapsto P_{n}\rho P_{n}+
\left(\mathrm{Tr}(I_{\mathcal{H}'}-P_{n})\rho\right)\tau_{n}$, the
monotonicity property of the relative entropy implies
$\bar{C}(\Phi_{n})\leq\bar{C}(\Phi)$ for all $n$. This and lemma 4
in \cite{Sh-4} imply the limit expressions in the second part of
the corollary.$\square$

The above corollary shows that the class of channels with finite
$\chi$-capacity is sufficiently close to the class of
$\textup{IF}$-channels. Nevertheless channels of this class
demonstrate many features of essential infinite dimensional
channels (see examples \ref{example-1} and \ref{example-2} below).

There exists a class of infinite dimensional channels with finite
$\chi$-capacity which is the most close to the class of
$\textup{IF}$-channels.

\begin{definition}\label{reqularity}
\textit{A channel
$\Phi:\mathfrak{S}(\mathcal{H})\mapsto\mathfrak{S}(\mathcal{H}')$
is called $\textup{CE}$-channel if the restriction of the quantum
entropy to the set $\overline{\Phi(\mathfrak{S}(\mathcal{H}))}$ is
continuous.}
\end{definition}

The results in \cite{Sh-4} provide different sufficient conditions
of the $\textup{CE}$-property of a channel (see proposition
\ref{sr-o-a-s} below). In particular, proposition 10 in \cite{Sh-4}
implies that the class of $\textup{CE}$-channels is closed under
tensor products: \textit{If
$\Phi:\mathfrak{S}(\mathcal{H})\mapsto\mathfrak{S}(\mathcal{H}')$
and
$\Psi:\mathfrak{S}(\mathcal{K})\mapsto\mathfrak{S}(\mathcal{K}')$
are $\textup{CE}$-channels then the channel
$\Phi\otimes\Psi:\mathfrak{S}(\mathcal{H}\otimes\mathcal{K})\mapsto\mathfrak{S}(\mathcal{H}'\otimes\mathcal{K}')$
is a $\textup{CE}$-channel.}

The $\chi$-capacity $\bar{C}(\Phi)$ of the channel $\Phi$ can be
defined as the least upper bound of the $\chi$-function of the
channel $\Phi$ defined by \cite{H-Sh-2}
\begin{equation*}
\chi_{\Phi}(\rho)=\!\sup_{\sum_{i}\pi _{i}\rho
_{i}=\rho}\sum_{i}\pi_{i}H(\Phi (\rho
_{i})\|\Phi(\rho))=\sup_{\mu\in\mathcal{P}_{\{\rho\}}}\int
\limits_{\mathfrak{S}(\mathcal{H})}H(\Phi(\sigma)\Vert\Phi(\rho))\mu(d\sigma),
\end{equation*}
where $\mathcal{P}_{\{\rho\}}$ is the set of all probability
measures on $\mathfrak{S}( \mathcal{H})$ with the barycenter
$\rho$.

It was shown in \cite{Sh-3} that the convex closure of the output
entropy is defined by
$$
\hat{H}_{\Phi}(\rho)= \inf_{\mu\in\mathcal{P}_{\{\rho\}}}
\int\limits_{\mathfrak{S}(\mathcal{H})}H(\Phi(\rho))\mu(d\rho)\leq+\infty.
$$
In contrast to the $\chi$-function the infinum over all measures
in $\mathcal{P}_{\{\rho\}}$ in the definition of the
$\hat{H}$-function does not coincide in general with the infinum
over all measures in $\mathcal{P}_{\{\rho\}}$ with finite support
\cite{Sh-3}.

The $\chi$-function and the $\hat{H}$-function of an arbitrary
channel are nonegative lower semicontinuous concave and convex
functions correspondingly \cite{Sh-3}. The following proposition
shows, in particular, that the $\chi$-function and the
$\hat{H}$-function of a $\textup{CE}$-channel have properties
similar to the properties of these functions of a finite
dimensional channel.

\begin{property}\label{chi-function}
\textit{Let $\Phi$ be channel with finite $\chi$-capacity. Then
\[
\chi _{\Phi}(\rho)\leq\bar{C}(\Phi)-H(\Phi (\rho )\Vert
\Omega(\Phi)),\quad \forall \rho \in \mathfrak{S}(\mathcal{H}).
\]
If $\Phi$ is a $\textup{CE}$-channel then the functions
$\chi_{\Phi}$ and $\hat{H}_{\Phi}$ are continuous on
$\mathfrak{S}(\mathcal{H})$. Moreover
$$
\chi_{\Phi}(\rho)=H(\Phi(\rho))-\hat{H}_{\Phi}(\rho)\quad and\quad
\hat{H}_{\Phi}(\rho)=\inf_{\sum_{i}\pi_{i}\rho_{i}=\rho}\sum_{i}\pi_{i}H(\Phi(\rho_{i})).
$$}
\end{property}

\textbf{Proof.} The inequality for the $\chi$-function follows from
corollary 1 in \cite{Sh-2}. The continuity assertion and the
representations for the $\chi$-function and the $\hat{H}$-function
are corollaries of proposition 7 and proposition 5 in
\cite{Sh-3}.$\square$

The following proposition shows the special role of the output
optimal average state.
\begin{property}\label{sr-o-a-s}
\textit{Let $\Phi$ be a channel with finite $\chi$-capacity.}

\textit{If $\mathrm{dc}(\Omega(\Phi))<1$ then
$\sup_{\rho\in\mathfrak{S}(\mathcal{H})}H(\Phi(\rho))<+\infty$ and
$\bar{C}(\Phi)\leq H(\Omega(\Phi))$.}

\textit{If $\mathrm{dc}(\Omega(\Phi))=0$ then $\Phi$ is a
$\textup{CE}$-channel.}
\end{property}

\textbf{Proof.} Suppose $\mathrm{dc}(\Omega(\Phi))<1$. By theorem 2E
in \cite{Sh-4} the entropy is bounded on the set
$\overline{\Phi(\mathfrak{S}(\mathcal{H}))}$. By theorem 1 for
arbitrary approximating sequence of ensembles
$\{\{\pi_{i}^{n},\rho_{i}^{n}\}\}_{n}$ the corresponding sequence
$\{\Phi(\bar{\rho}_{n})\}_{n}$ of images of their average states
$H$-converges to the state $\Omega(\Phi)$. By proposition 2 in
\cite{Sh-4} the condition $\mathrm{dc}(\Omega(\Phi))<1$ implies
$$
\bar{C}(\Phi)=\lim_{n\rightarrow+\infty}\chi_{\Phi}(\{\pi_{i}^{n},\rho_{i}^{n}\})\leq
\lim_{n\rightarrow+\infty}H(\Phi(\bar{\rho}_{n}))=H(\Omega(\Phi)).
$$

Suppose $\mathrm{dc}(\Omega(\Phi))=0$. By theorem 2E in
\cite{Sh-4} the entropy is continuous on the set
$\overline{\Phi(\mathfrak{S}(\mathcal{H}))}$. $\square$

The assertions of the above proposition are illustrated by the
below example \ref{example-1}, in which the family of channels
with different decrease coefficient of the output optimal average
state is considered.

According to \cite{H-Sh-2} a measure $\mu_{*}$ in
$\mathcal{P}(\mathrm{Ext}\mathfrak{S}(\mathcal{H}))$, at which the
supremum in definition (\ref{ccap-2}) is achieved, is called
\textit{optimal measure (optimal generalized ensemble)} for the
channel $\Phi$.

The notion of an optimal measure is a generalization of the notion
of an optimal ensemble for a finite dimensional channel. In
\cite{Sch-West-1} it is shown that each optimal ensemble is
characterized by the maximal distance property. The first part of
the following theorem generalizes this observation to the case of
infinite dimensional channel.

\begin{theorem}\label{o-m} \textit{Let
$\Phi:\mathfrak{S}(\mathcal{H})\mapsto\mathfrak{S}(\mathcal{H}')$
be a channel with finite $\chi$-capacity.}

\textit{If there exists an optimal measure $\mu_{*}$ for the
channel $\Phi$ then its barycenter $\bar{\rho}(\mu_{*})$ is an
input optimal average state for the channel $\Phi$ and the
following "maximal distance property" holds
$$
H(\Phi(\rho)\|\Omega(\Phi))=\bar{C}(\Phi)\;for\;
\mu_{*}-almost\;all\;\rho \;in\; \mathfrak{S}(\mathcal{H}).
$$}\vspace{-10pt}

\textit{If there exists an input optimal average state $\rho_{*}$
for the channel $\Phi$ and the set $\Phi(\mathfrak{S}(\mathcal{H}))$
is regular then there exists an optimal measure $\mu_{*}$ for the
channel $\Phi$ such that $\bar{\rho}(\mu_{*})=\rho_{*}$.}
\end{theorem}

\textbf{Proof.} Since every probability measure can be weakly
approximated by a sequence of measures with finite support lower
semicontinuity of the functional $\chi_{\Phi}$ implies that the
barycenter $\bar{\rho}(\mu_{*})$ of any optimal measure $\mu_{*}$
for the channel $\Phi$ is an input optimal average state for the
channel $\Phi$. The maximal distance property follows from the
definition of an optimal measure and theorem 1.

Suppose the set $\Phi(\mathfrak{S}(\mathcal{H}))$ is regular and
$\{\mu_{n}=\{\pi^{n}_{i}, \rho^{n}_{i}\}\}_{n}$ is an
approximating sequence of ensembles for the channel $\Phi$ such
that the corresponding sequence $\{\bar{\rho}(\mu_{n})\}_{n}$
converges to the state $\rho_{*}$. By convexity and lower
semicontinuity of the relative entropy we may assume that each
ensemble in this sequence consists of pure states. Since the
sequence $\{\bar{\rho}(\mu_{n})\}_{n}$ is relatively compact the
sequence $\{\mu_{n}\}_{n}$ is relatively weakly compact by
proposition 2 in \cite{H-Sh-2} and hence it contains weakly
converging subsequence. So, we may assume that the sequence
$\{\mu_{n}\}_{n}$ weakly converges to a particular measure
$\mu_{*}$ supported by pure states due to theorem 6.1 in
\cite{Par}.

For given $n$ let $\nu_{n}=\mu_{n}\circ\Phi^{-1}$ be the image of
the measure $\mu_{n}$ under the mapping $\Phi$, so that
$\nu_{n}=\{\pi^{n}_{i}, \Phi(\rho^{n}_{i})\}$. It is easy to see
that $\{\nu_{n}\}_{n}$ is an approximating sequence of ensembles
for the set $\overline{\Phi(\mathfrak{S}(\mathcal{H}))}$. Since
this set is regular by the condition the arguments from the proof
of theorem 3 and lemma 5 in \cite{Sh-4} imply existence of a
subsequence $\{\nu_{n_{k}}\}_{k}$ weakly converging to an optimal
measure $\nu_{*}$ for the set
$\overline{\Phi(\mathfrak{S}(\mathcal{H}))}$. But
$\mu_{n_{k}}\rightarrow\mu_{*}$ in the weak topology implies
$\mu_{n_{k}}\circ\Phi^{-1}\rightarrow\mu_{*}\circ\Phi^{-1}$ in the
weak topology. So, we obtain $\mu_{*}\circ\Phi^{-1}=\nu_{*}$. Thus
$\mu_{*}$ is an optimal measure for the channel $\Phi$. $\square$

Theorem \ref{o-m} and theorem 2E in \cite{Sh-4} imply the
following sufficient condition of existence of an optimal measure.

\begin{corollary}\label{o-m-s-c}
\textit{Existence of an input optimal average state  for the
channel $\Phi$ implies existence of an optimal measure for the
channel $\Phi$ provided one of the following conditions holds:}
\begin{itemize}
  \item \textit{$\mathrm{dc}(\Omega(\Phi))<1$;}
  \item \textit{the channel
$\Phi$ is a $\textup{CE}$-channel.}
\end{itemize}

\end{corollary}

\begin{remark}\label{on o-m-s-c}
The conditions in theorem \ref{o-m} and in corollary \ref{o-m-s-c}
are essential as it is shown by the examples of channels with
finite $\chi$-capacity for which there exist no optimal measures
(examples \ref{example-1}C and \ref{example-2}). The meaning of
these conditions are considered in remark \ref{on i-o-a-s}
below.$\square$
\end{remark}

The above general observations are illustrated by the following
family of channels, depending on a sequence of positive numbers.

\begin{example}\label{example-1} Let $\{|n\rangle\}_{n\in\mathbb{N}\cup\{0\}}$ be an orthonormal basis in
$\mathcal{H}'$ and $\{q_{n}\}_{n\in\mathbb{N}}$ be a sequence of
numbers in $(0,1]$, converging to zero. Consider the set
$\mathcal{S}^{1}_{\{q_{n}\}}$ consisting of pure states
$$
\sigma_{k}=(1-q_{|k|})|0\rangle\langle
0|+q_{|k|}||k|\rangle\langle
|k||+\mathrm{sign}(k)\sqrt{(1-q_{|k|})q_{|k|}}(|0\rangle\langle
|k||+||k|\rangle\langle 0|),
$$
indexed by the set $\mathbb{Z}\setminus\{0\}$. This set can be
considered as a sequence of states converging to the state
$|0\rangle\langle 0|$ and its properties were explored in
subsection 5.1 in \cite{Sh-4}.

Consider the channel
\begin{equation*}
\Phi_{\{q_{n}\}}(\rho)=\sum_{k}\langle k|\rho|k \rangle\sigma_{k},
\end{equation*}
where $\{|k\rangle\}_{k\in\mathbb{Z}\setminus\{0\}}$ be an
orthonormal basis in $\mathcal{H}$.

This channel is entanglement-breaking \cite{R},\cite{H-Sh-W} and
$\overline{\Phi_{\{q_{n}\}}(\mathfrak{S}(\mathcal{H}))}=\overline{\mathrm{co}}(\mathcal{S}^{1}_{\{q_{n}\}})$
is a compact subset of $\mathfrak{S}(\mathcal{H}')$. Since the
states in the sequence $\mathcal{S}^{1}_{\{q_{n}\}}$ are pure
$\bar{C}(\Phi_{\{q_{n}\}})=\sup_{\rho\in\mathfrak{S}(\mathcal{H})}H(\Phi_{\{q_{n}\}}(\rho))$.
Let $\lambda^{*}_{\{q_{n}\}}$ be the infinum of the set
$\left\{\lambda:\;\sum_{n}\exp\left(-\frac{\lambda}{q_{n}}\right)<+\infty\right\}$
if it is not empty and $\lambda^{*}_{\{q_{n}\}}=+\infty$
otherwise. Let
$$
F_{\{q_{n}\}}(\lambda)=\sum_{k=1}^{+\infty}\left(\frac{1}{q_{k}}-1\right)\exp\left(-\frac{\lambda}{q_{k}}\right)
$$
be a decreasing function on $[\lambda^{*}_{\{q_{n}\}},+\infty)$ with
the range $(0,+\infty]$.

We will show that:

\begin{enumerate} [A)]

\item\textit{ If $\lambda^{*}_{\{q_{n}\}}=0$ then
\begin{itemize}
  \item $\Phi_{\{q_{n}\}}$ is a $\textup{CE}$-channel;
  \item there exist the unique optimal measure and the unique input optimal average
        state for the channel $\Phi_{\{q_{n}\}}$;
  \item
  $\Omega(\Phi_{\{q_{n}\}})\in\Phi_{\{q_{n}\}}(\mathfrak{S}(\mathcal{H}))$,\\
  $\mathrm{dc}(\Omega(\Phi_{\{q_{n}\}}))=0$ and
  $\bar{C}(\Phi_{\{q_{n}\}})=H(\Omega(\Phi_{\{q_{n}\}}))$;
\end{itemize}}

\item \textit{If $0<\lambda^{*}_{\{q_{n}\}}<+\infty$ and $F_{\{q_{n}\}}(\lambda^{*}_{\{q_{n}\}})\geq 1$ then
\begin{itemize}
  \item $\bar{C}(\Phi_{\{q_{n}\}})<+\infty$, but $\Phi_{\{q_{n}\}}$ is not a $\textup{CE}$-channel;
  \item there exist the unique optimal measure and the unique input optimal average
        state for the channel $\Phi_{\{q_{n}\}}$;
  \item
  $\Omega(\Phi_{\{q_{n}\}})\in\Phi_{\{q_{n}\}}(\mathfrak{S}(\mathcal{H}))$,\\
  $0<\mathrm{dc}(\Omega(\Phi_{\{q_{n}\}}))\leq 1$
  \footnote{In this case $\mathrm{dc}(\Omega(\Phi_{\{q_{n}\}}))=1$ if and only if
  $F_{\{q_{n}\}}(\lambda^{*}_{\{q_{n}\}})=1.$} and $\bar{C}(\Phi_{\{q_{n}\}})=H(\Omega(\Phi_{\{q_{n}\}}))$;
\end{itemize}}

\item \textit{If $0<\lambda^{*}_{\{q_{n}\}}<+\infty$ and $F_{\{q_{n}\}}(\lambda^{*}_{\{q_{n}\}})< 1$ then
\begin{itemize}
  \item $\bar{C}(\Phi_{\{q_{n}\}})<+\infty$, but $\Phi_{\{q_{n}\}}$ is not a $\textup{CE}$-channel;
  \item there exist no optimal measure and no input optimal average
        state for the channel $\Phi_{\{q_{n}\}}$;
  \item
  $\Omega(\Phi_{\{q_{n}\}})\in\overline{\Phi_{\{q_{n}\}}(\mathfrak{S}(\mathcal{H}))}\setminus
  \Phi_{\{q_{n}\}}(\mathfrak{S}(\mathcal{H}))$,\\ $\mathrm{dc}(\Omega(\Phi_{\{q_{n}\}}))=1$
  and
  $\bar{C}(\Phi_{\{q_{n}\}})>H(\Omega(\Phi_{\{q_{n}\}}))$;
\end{itemize}
}
\item \textit{If $\lambda^{*}_{\{q_{n}\}}=+\infty$ then
$\bar{C}(\Phi_{\{q_{n}\}})=+\infty$.}
\end{enumerate}

\textit{In the cases A and B the $\chi$-capacity, the unique
optimal measure, the unique input and output optimal average
states are defined by the following expressions (correspondingly):
$$
\bar{C}\left(\Phi_{\{q_{n}\}}\right)
=\lambda_{\{q_{n}\}}^{1}-\log\pi_{\{q_{n}\}}^{1},\quad
\mu_{*}=\left\{\frac{\pi_{\{q_{n}\}}^{1}}{2q_{|k|}}\exp\left(-\frac{\lambda_{n}}{q_{|k|}}\right);|k\rangle\langle
k|\right\}_{k\in\mathbb{Z}\setminus\{0\}},
$$
$$
\bar{\rho}(\mu_{*})=
\sum_{k\in\mathbb{Z}\setminus\{0\}}\frac{\pi_{\{q_{n}\}}^{1}}{2q_{|k|}}
\exp\left(-\frac{\lambda_{\{q_{n}\}}^{1}}{q_{|k|}}\right)|k\rangle\langle
k|,
$$
$$
\Omega\left(\Phi_{\{q_{n}\}}\right)=\pi_{\{q_{n}\}}^{1}|0\rangle\langle
0|+\pi_{\{q_{n}\}}^{1}\sum_{n=1}^{+\infty}
\exp\left(-\frac{\lambda_{\{q_{n}\}}^{1}}{q_{n}}\right)|n\rangle\langle
n|,
$$
where $\lambda_{\{q_{n}\}}^{1}$ is the unique solution of the
equation $F_{\{q_{n}\}}(\lambda)=1$ and
$$
\pi_{\{q_{n}\}}^{1}=\left(1+\sum_{n=1}^{+\infty}
\exp\left(-\frac{\lambda_{\{q_{n}\}}^{1}}{q_{n}}\right)\right)^{-1}=\left(\sum_{n=1}^{+\infty}\frac{1}{q_{n}}
\exp\left(-\frac{\lambda_{\{q_{n}\}}^{1}}{q_{n}}\right)\right)^{-1}\in[0,1].
$$}

\textit{In the case C the $\chi$-capacity and the output optimal
average state are defined by the following expressions
(correspondingly):
$$
\bar{C}\left(\Phi_{\{q_{n}\}}\right)
=\lambda_{\{q_{n}\}}^{*}-\log\pi_{\{q_{n}\}}^{*},
$$
$$
\Omega\left(\Phi_{\{q_{n}\}}\right)=\pi_{\{q_{n}\}}^{*}|0\rangle\langle
0|+\pi_{\{q_{n}\}}^{*}\sum_{n=1}^{+\infty}
\exp\left(-\frac{\lambda_{\{q_{n}\}}^{*}}{q_{n}}\right)|n\rangle\langle
n|,
$$
where
$$
\pi_{\{q_{n}\}}^{*}=\left(1+\sum_{n=1}^{+\infty}
\exp\left(-\frac{\lambda_{\{q_{n}\}}^{*}}{q_{n}}\right)\right)^{-1}\in[0,1].
$$}

If $\lambda^{*}_{\{q_{n}\}}=+\infty$ then
$\bar{C}(\Phi_{\{q_{n}\}})=\bar{C}(\mathcal{S}^{1}_{\{q_{n}\}})=+\infty$
by the observation in subsection 5.1 in \cite{Sh-4}.

Suppose $\lambda^{*}_{\{q_{n}\}}<+\infty$. Without loss of
generality we may assume that the sequence $\{q_{n}\}$ is
nonincreasing. Let
\begin{equation}\label{H-operator}
H=0|0\rangle\langle
0|+\sum_{n=1}^{+\infty}q_{n}^{-1}|n\rangle\langle n|
\end{equation}
be a $\mathfrak{H}$-operator in $\mathcal{H}'$ so that
$\mathrm{ic}(H)=\lambda^{*}_{\{q_{n}\}}$ (see section 3 in
\cite{Sh-4}). It is easy to see that
$\mathcal{S}^{\varepsilon}_{\{q_{n}\}}\subseteq \mathcal{K}_{H,1}$
and hence proposition 1b in \cite{Sh-4} implies
$\bar{C}(\Phi_{\{q_{n}\}})=\bar{C}(\mathcal{S}^{1}_{\{q_{n}\}})\leq
\bar{C}(\mathcal{K}_{H,1})=\sup_{\rho\in\mathcal{K}_{H,1}}H(\rho)<+\infty$.
We will show that
$\bar{C}(\mathcal{S}^{1}_{\{q_{n}\}})=\bar{C}(\mathcal{K}_{H,1})$.
In the proof of proposition 1a in \cite{Sh-4} the sequence
$\{\rho_{n}\}$ of states defined by formula (9) was constructed
and it was shown that
$\lim_{n\rightarrow+\infty}H(\rho_{n})=\sup_{\rho\in\mathcal{K}_{H,1}}H(\rho)$.
For the $\mathfrak{H}$-operator $H$ defined by (\ref{H-operator})
the states of the above sequence have the form
\begin{equation}\label{rho-n}
\rho_{n}=\left(1+\sum_{k=1}^{n}\exp\left(-\frac{\lambda_{n}}{q_{k}}\right)\right)^{-1}
\left(|0\rangle\langle
0|+\sum_{k=1}^{n}\exp\left(-\frac{\lambda_{n}}{q_{k}}\right)|k\rangle\langle
k|\right)
\end{equation}
for sufficiently large $n\in\mathbb{N}$, where $\lambda_{n}$ is
the unique solution of the equation
\begin{equation}\label{lambda-n-eq}
1+\sum_{k=1}^{n}\exp\left(-\frac{\lambda}{q_{k}}\right)=
\sum_{k=1}^{n}\frac{1}{q_{k}}\exp\left(-\frac{\lambda}{q_{k}}\right).
\end{equation}

The sequence of states defined by (\ref{rho-n}) lies in
$\overline{\mathrm{co}}(\mathcal{S}^{1}_{\{q_{n}\}})$. Indeed, for
given $n$ let
$$
\pi_{k}^{n}=\left(\sum_{k=1}^{n}\frac{2}{q_{|k|}}
\exp\left(-\frac{\lambda_{n}}{q_{|k|}}\right)\right)^{-1}
\frac{1}{q_{|k|}}\exp\left(-\frac{\lambda_{n}}{q_{|k|}}\right),\;k\in\mathbb{Z}\setminus\{0\},
$$
be a probability distribution. By using equality (\ref{lambda-n-eq})
with $\lambda=\lambda_{n}$ it is easy to see that
$\sum_{k}\pi_{k}^{n}\sigma_{k}=\rho_{n}$ and hence
$\rho_{n}\in\overline{\mathrm{co}}(\mathcal{S}^{1}_{\{q_{n}\}})$.
This, proposition 1b and theorem 2C in \cite{Sh-4} imply
\begin{equation}\label{cap-coincidence}
\bar{C}(\Phi_{\{q_{n}\}})=\bar{C}(\mathcal{S}^{1}_{\{q_{n}\}})=
\bar{C}(\mathcal{K}_{H,1})=\lambda^{*}-\log\pi^{*}
\end{equation}
and
\begin{equation}\label{oas-coincidence}
\Omega(\Phi_{\{q_{n}\}})=\Omega(\mathcal{S}^{1}_{\{q_{n}\}})=
\Omega(\mathcal{K}_{H,1})=\pi^{*}|0\rangle\langle
0|+\pi^{*}\sum_{n=1}^{+\infty}
\exp\left(-\frac{\lambda^{*}}{q_{n}}\right)|n\rangle\langle n|,
\end{equation}
where $\pi^{*}=1+\displaystyle\sum_{n=1}^{+\infty}
\exp\left(-\frac{\lambda^{*}}{q_{n}}\right)$ and $\lambda^{*}$ is
the unique solution of the equation
\begin{equation}\label{lambda-*-eq}
1+\sum_{k=1}^{+\infty}\exp\left(-\frac{\lambda}{q_{k}}\right)=
\sum_{k=1}^{+\infty}\frac{1}{q_{k}}\exp\left(-\frac{\lambda}{q_{k}}\right)
\end{equation}
if
\begin{equation}\label{m-ineq}
h_{*}(H)=\frac{\displaystyle\mathrm{Tr}H\exp(-\mathrm{ic}(H)H)}{\displaystyle\mathrm{Tr}\exp(-\mathrm{ic}(H)H)}=
\frac{\displaystyle\sum_{k=1}^{+\infty}\frac{1}{q_{k}}\exp\left(-\frac{\lambda^{*}_{\{q_{n}\}}}{q_{k}}\right)}
{\displaystyle
1+\sum_{k=1}^{+\infty}\exp\left(-\frac{\lambda^{*}_{\{q_{n}\}}}{q_{k}}\right)}\geq
h=1
\end{equation}
and $\lambda^{*}=\lambda^{*}_{\{q_{n}\}}$ otherwise. It is easy to
see that (\ref{lambda-*-eq}) is equivalent to the equation
$F_{\{q_{n}\}}(\lambda)=1$ while (\ref{m-ineq}) means the
inequality $F_{\{q_{n}\}}(\lambda^{*}_{\{q_{n}\}})\geq 1$.

The assertion concerning existence of the unique optimal measure
$\mu_{*}$ for the channel $\Phi_{\{q_{n}\}}$ in the cases A and B
and the expression for this measure follows from the observation
in subsection 5.1 in \cite{Sh-4} (with $\varepsilon=1$). The
barycenter of the measure $\mu_{*}$ is the unique input optimal
average state for the channel $\Phi_{\{q_{n}\}}$. Uniqueness of
the input optimal average state follows from uniqueness of the
optimal measure, since proposition 1b in \cite{Sh-4} and
(\ref{oas-coincidence}) imply regularity of the set
$\Phi_{\{q_{n}\}}(\mathfrak{S}(\mathcal{H}))\subseteq
\mathcal{K}_{H,1}$ in the cases A and B and hence theorem 2 shows
that each input optimal average state is a barycenter of at least
one optimal measure.

In the cases A and B the state $\Omega(\Phi_{\{q_{n}\}})$ is an
image of the barycenter of the optimal measure and hence it lies
in $\Phi(\mathfrak{S}(\mathcal{H}))$. To prove nonexistence of an
input optimal average state for the channel $\Phi_{\{q_{n}\}}$ in
the case C it is sufficient to show that in this case the state
$\Omega(\Phi_{\{q_{n}\}})$ does not lie in
$\Phi(\mathfrak{S}(\mathcal{H}))$.

The set $\Phi(\mathfrak{S}(\mathcal{H}))$ is a $\sigma$-convex
hull of the set
$\mathcal{S}^{1}_{\{q_{n}\}}=\{\sigma_{k}\}_{k\in\mathbb{Z}\setminus\{0\}}$,
so that the assumption
$\Omega(\Phi_{\{q_{n}\}})\in\Phi(\mathfrak{S}(\mathcal{H}))$
implies existence of a probability distribution
$\{\pi_{k}\}_{k\in\mathbb{Z}\setminus\{0\}}$ such that
$\Omega(\Phi_{\{q_{n}\}})=\sum_{k\in\mathbb{Z}\setminus\{0\}}\pi_{k}\sigma_{k}$.
By using the expression for the state $\Omega(\Phi_{\{q_{n}\}})$
in the case C we obtain from this decomposition that
$$
q_{n}(\pi_{-n}+\pi_{n})=\frac{\displaystyle\exp\left(-\frac{\lambda^{*}_{\{q_{n}\}}}{q_{n}}\right)}
{\displaystyle
1+\sum_{k=1}^{+\infty}\exp\left(-\frac{\lambda^{*}_{\{q_{n}\}}}{q_{k}}\right)},\quad
\forall n\in\mathbb{N},
$$
and hence
$$
\sum_{k=1}^{+\infty}\frac{1}{q_{k}}\exp\left(-\frac{\lambda^{*}_{\{q_{n}\}}}{q_{k}}\right)=
1+\sum_{k=1}^{+\infty}\exp\left(-\frac{\lambda^{*}_{\{q_{n}\}}}{q_{k}}\right).
$$
But this equality means that the equality holds in inequality (16)
in \cite{Sh-4} for the $\mathfrak{H}$-operator $H$ defined by
(\ref{H-operator}). By the observation in \cite{Sh-4} equality in
inequality (16) means inequality $h_{*}(H)\geq h=1$ equivalent to
the inequality $F_{\{q_{n}\}}(\lambda^{*}_{\{q_{n}\}})\geq 1$,
which contradicts to the definition of the case C.

If $\lambda^{*}_{\{q_{n}\}}=\mathrm{ic}(H)=0$ then the above
observation implies $\mathrm{dc}(\Omega(\Phi_{\{q_{n}\}})=0$ and
by proposition \ref{sr-o-a-s} $\Phi_{\{q_{n}\}}$ is a
$\mathrm{CE}$-channel.

Suppose the entropy is continuous on the set
$\overline{\mathrm{co}}(\mathcal{S}^{1}_{\{q_{n}\}})$. Consider
the sequence of states
$$
\rho_{n}=\left(1-\sum_{k=1}^{n}q_{k}\pi_{k}\right)|0\rangle\langle
0|+\sum_{k=1}^{n}q_{k}\pi_{k}|k\rangle\langle k|,
$$
where
$\{\pi_{k}=q_{k}^{-1}(\sum_{k=1}^{n}q_{k}^{-1})^{-1}\}_{k=1}^{n}$
is a probability distribution. It is easy to see that this
sequence lies in $\mathrm{co}(\mathcal{S}^{1}_{\{q_{n}\}})$ and
converges to the pure state $|0\rangle\langle 0|$. By the
continuity assumption $\lim_{n\rightarrow+\infty}H(\rho_{n})=0$,
which implies
\begin{equation}\label{lim-exp}
\lim_{n\rightarrow+\infty}n(q_{k}\pi_{k}(-\log(q_{k}\pi_{k})))=
\lim_{n\rightarrow+\infty}nf\left(\sum_{k=1}^{n}q_{k}^{-1}\right)=0,
\end{equation}
where $f(x)=\log x/x$. Since the function $f(x)$ is decreasing for
large $x$ the obvious inequality $\sum_{k=1}^{n}q_{k}^{-1}\leq
nq_{n}^{-1}$ and (\ref{lim-exp}) imply
$\lim_{n\rightarrow+\infty}\nu_{n}=0$, where
$\nu_{n}=nf(nq_{n}^{-1})=q_{n}\log(nq_{n}^{-1})$. Hence for
arbitrary $\lambda>0$ we have
$$
\left(\frac{q_{n}}{n}\right)^{\frac{\lambda}{\nu_{n}}}=\exp\left(-\frac{\lambda}{q_{n}}\right),
$$
which implies $\lambda^{*}_{\{q_{n}\}}=0$. $\square$

The most interesting case of the above example is the case C, in
which the channel $\Phi_{\{q_{n}\}}$ demonstrates essential
infinite dimensional features.  The example of a sequence
$\{q_{n}\}$ corresponding to this case can be found in
\cite{Sh-4}.

Example \ref{example-1} can be generalized by considering the set
$\mathcal{S}^{\varepsilon}_{\{q_{n}\}}$ with arbitrary
$\varepsilon$ in $[0,1]$ instead of $\mathcal{S}^{1}_{\{q_{n}\}}$
\cite{Sh-4}.
\end{example}

\section{On extension to nonnormal states}

Let $\widehat{\mathfrak{S}}(\mathcal{H})$ be the set of all
normalized positive functionals on $\mathfrak{B}(\mathcal{H})$, so
that
$\widehat{\mathfrak{S}}(\mathcal{H})\subset(\mathfrak{B}(\mathcal{H}))^{*}$.
It is known that $\widehat{\mathfrak{S}}(\mathcal{H})$ is compact
in the $*$-weak topology and that $\mathfrak{S}(\mathcal{H})$ can
be considered as a $*$-weak dense subset of
$\widehat{\mathfrak{S}}(\mathcal{H})$ \cite{B&R}. So, it is
natural to explore a possibility to extend an arbitrary channel
from $\mathfrak{S}(\mathcal{H})$ to
$\widehat{\mathfrak{S}}(\mathcal{H})$. The following proposition
provides a simple characterization of the class of channels, which
has natural extension to $\widehat{\mathfrak{S}}(\mathcal{H})$.

\begin{property}\label{compact output}\footnote{This simple observation seems to be a corollary of
a general result in the functional analysis. The author would be
grateful for any references.} \textit{A channel
$\Phi:\mathfrak{S}(\mathcal{H})\mapsto\mathfrak{S}(\mathcal{H}')$
can be extended to the mapping
$\widehat{\Phi}:\widehat{\mathfrak{S}}(\mathcal{H})\mapsto\mathfrak{S}(\mathcal{H}')$
continuous with respect to the $*$-weak topology on the set
$\widehat{\mathfrak{S}}(\mathcal{H})$ and the trace norm topology
on the set $\mathfrak{S}(\mathcal{H}')$ if and only if the set
$\Phi(\mathfrak{S}(\mathcal{H}))$ is relatively compact.}

\textit{If the above extension $\widehat{\Phi}$ exists then
$\overline{\Phi(\mathfrak{S}(\mathcal{H}))}=\widehat{\Phi}(\widehat{\mathfrak{S}}(\mathcal{H}))$.}

\end{property}

\textbf{Proof.} If $\widehat{\Phi}$ is the above extension of the
channel $\Phi$ then
$\widehat{\Phi}(\widehat{\mathfrak{S}}(\mathcal{H}))$ is compact
as an image of a compact set under a continuous mapping. Since
$\mathfrak{S}(\mathcal{H})$ is a $*$-weak dense subset of
$\widehat{\mathfrak{S}}(\mathcal{H})$ we have
$\overline{\Phi(\mathfrak{S}(\mathcal{H}))}=\widehat{\Phi}(\widehat{\mathfrak{S}}(\mathcal{H}))$.

Suppose the set $\overline{\Phi(\mathfrak{S}(\mathcal{H}))}$ is
compact. Since
$\mathfrak{B}(\mathcal{H})^{*}=\mathfrak{T}(\mathcal{H})^{**}$
there exists the linear mapping
$\Phi^{**}:\mathfrak{B}(\mathcal{H})^{*}\mapsto\mathfrak{B}(\mathcal{H}')^{*}$
continuous with respect to the $*$-weak topologies on the both
spaces and such that
$\Phi^{**}|_{\mathfrak{S}(\mathcal{H})}=\Phi$. By using the
compactness argument and $*$-weak density of the set
$\mathfrak{S}(\mathcal{H})$ in
$\widehat{\mathfrak{S}}(\mathcal{H})$ it is easy to show that
$\Phi^{**}(\widehat{\mathfrak{S}}(\mathcal{H}))\subseteq\overline{\Phi(\mathfrak{S}(\mathcal{H}))}$.
Since each one-to-one continuous mapping from compact topological
space onto Hausdorff topological space is a hemeomorphism
\cite{K&F} we can conclude that the identity mapping from
$\overline{\Phi(\mathfrak{S}(\mathcal{H}))}$ with the trace norm
topology onto itself with the $*$-weak topology has a continuous
converse. This and the previous observation imply that $\Phi^{**}$
is continuous with respect to the $*$-weak topology on
$\widehat{\mathfrak{S}}(\mathcal{H})$ and the trace norm topology
on $\mathfrak{S}(\mathcal{H}')$. So, the mapping
$\Phi^{**}|_{\widehat{\mathfrak{S}}(\mathcal{H})}$ has all the
properties of the extension $\widehat{\Phi}$, stated in the
proposition. $\square$

\begin{definition}\label{ext-def}
\textit{For an arbitrary channel $\Phi$ with relatively compact
output the mapping $\widehat{\Phi}$ introduced in proposition
\ref{compact output} is called its channel extension.}
\end{definition}

Proposition \ref{compact output} and theorem \ref{o-a-s} imply the
following observation.
\begin{corollary}\label{c-f-c-ext}
\textit{If $\Phi$ is a channel with finite $\chi$-capacity then it
has the channel extension $\widehat{\Phi}$.}
\end{corollary}

By this corollary for given channel $\Phi$ with finite
$\chi$-capacity it is possible to consider the entropic
characteristics of its channel extension $\widehat{\Phi}$ such as
the minimal output entropy $H_{\mathrm{min}}(\widehat{\Phi})=
\inf_{\hat{\rho}\in\widehat{\mathfrak{S}}(\mathcal{H})}H(\widehat{\Phi}(\hat{\rho}))$
and the $\chi$-capacity
\begin{equation}
\bar{C}(\widehat{\Phi})=\sup_{\{\pi_{i},\hat{\rho}_{i}\}}\sum_{i}\pi_{i}H(\widehat{\Phi}(\hat{\rho}
_{i})\Vert\widehat{\Phi}(\textstyle\sum_{j}\pi_{j}\hat{\rho}_{j})),
\label{ccap-ext}
\end{equation}
where the supremum is over all finite ensembles
$\{\pi_{i},\hat{\rho}_{i}\}$ of states in
$\widehat{\mathfrak{S}}(\mathcal{H})$. By proposition \ref{compact
output} and theorem 2B in \cite{Sh-4} we have
\begin{equation}\label{cap-eq}
\bar{C}(\widehat{\Phi})=\bar{C}(\overline{\Phi(\mathfrak{S}(\mathcal{H}))})=
\bar{C}(\Phi(\mathfrak{S}(\mathcal{H})))=\bar{C}(\Phi).
\end{equation}

This means that one can not increase the $\chi$-capacity by using
nonnormal states.

In the same way as for the initial channel $\Phi$ we may define
the notions of an approximating sequence of ensembles and of an
input optimal average state for the channel extension
$\widehat{\Phi}$. In contrast to the case of the initial channel
$\Phi$ compactness of the set
$\widehat{\mathfrak{S}}(\mathcal{H})$ guarantees existence of at
least one input optimal average state for the channel extension
$\widehat{\Phi}$.  By using lower semicontinuity of the relative
entropy and (\ref{cap-eq}) it is possible to show that each
\textit{normal} input optimal average state for the channel
extension $\widehat{\Phi}$ is an input optimal average state for
the initial channel $\Phi$ and vice versa. For the channel
extension $\widehat{\Phi}$ it is possible to prove the analog of
theorem \ref{o-a-s}, in particular, to show that the image of any
input optimal average state for the channel extension
$\widehat{\Phi}$ coincides with the output optimal average state
$\Omega(\Phi)$.

In contrast to the $\chi$-capacity the minimal output entropies
for the channel $\Phi$ and for its extension $\widehat{\Phi}$ may
be different as it follows from the below simple example. Let
$\{\rho_{n}\}_{n=1}^{+\infty}$ be a sequence of states in
$\mathfrak{S}(\mathcal{H}')$ with infinite entropy, converging to
pure state $\rho_{0}$ and such that
$\bar{C}(\{\rho_{n}\}_{n=1}^{+\infty})<+\infty$. Consider the
channel $\Phi(\rho)=\sum_{n=1}^{+\infty}\langle
n|\rho|n\rangle\rho_{n}$, where $\{|n\rangle\}$ is some
orthonormal basis of $\mathcal{H}$. Since an arbitrary state in
$\Phi(\mathfrak{S}(\mathcal{H}))$ majorizes at least one trace
class operator with infinite entropy it has infinite entropy as
well \cite{W}. So we have $H_{\mathrm{min}}(\Phi)=+\infty$. But by
proposition \ref{compact output} we have
$\widehat{\Phi}(\widehat{\mathfrak{S}}(\mathcal{H}))=\overline{\mathrm{co}}(\{\rho_{n}\}_{n=1}^{+\infty})$
and hence there exists a state $\widehat{\rho}_{0}$ in
$\widehat{\mathfrak{S}}(\mathcal{H})$ such that
$\widehat{\Phi}(\widehat{\rho}_{0})=\rho_{0}$. This implies
$H_{\mathrm{min}}(\widehat{\Phi})=0$.

Let $\widehat{\mathcal{P}}$ be the set of all regular Borel
measures on the set $\widehat{\mathfrak{S}}(\mathcal{H})$ endowed
with the $*$-weak topology. In contrast to the set $\mathcal{P}$
the set $\widehat{\mathcal{P}}$ is not metrizable but it is
compact in the weak topology \cite{Alf}.

In the same way \footnote{The only difference in the argumentation
is a necessity to use proposition I.2.3 in \cite{Alf} instead of
lemma 1 in \cite{H-Sh-2} since the set
$\widehat{\mathfrak{S}}(\mathcal{H})$ is not complete separable
metric space, but it is compact.} as in \cite{H-Sh-2} it is
possible to show that
$$
\bar{C}(\widehat{\Phi})=\sup_{\hat{\pi}\in\widehat{\mathcal{P}}}
\int\limits_{\widehat{\mathfrak{S}}(\mathcal{H})}
H(\widehat{\Phi}(\hat{\rho})\Vert\widehat{\Phi}(\bar{\rho}(\hat{\pi})))\hat{\pi}(d\hat{\rho})
$$
where the supremum is over all probability measures on
$\widehat{\mathfrak{S}}(\mathcal{H})$ and $\bar{\rho}(\hat{\pi})$
is a barycenter of the measure $\hat{\pi}$. The measure
$\hat{\pi}_{*}$ at which the above supremum is achieved (if it
exists) is called \textit{optimal} for the channel extension
$\widehat{\Phi}$.

By using lower semicontinuity of the relative entropy, equality
(\ref{cap-eq}) and theorem \ref{o-a-s} it is possible to prove
that any optimal measure $\hat{\pi}$ for the channel extension
$\widehat{\Phi}$ has the generalized maximal distance property:
$$
H(\widehat{\Phi}(\hat{\rho})\|\Omega(\Phi))=\bar{C}(\widehat{\Phi})
\quad \textup{for}\;\hat{\pi}_{*}\textup{-almost}\;\textup{all}\;
\hat{\rho}\;\textup{in}\;\widehat{\mathfrak{S}}(\mathcal{H}).
$$

The following proposition shows, in particular, that the condition
of existence of an optimal measure for the channel extension
$\widehat{\Phi}$ is substantially weaker than the condition of
existence of an optimal measure for the initial channel $\Phi$.

\begin{property}\label{o-m-ext}
\textit{Let
$\Phi:\mathfrak{S}(\mathcal{H})\mapsto\mathfrak{S}(\mathcal{H}')$
be a channel with finite $\chi$-capacity and
$\widehat{\Phi}:\widehat{\mathfrak{S}}(\mathcal{H})\mapsto\mathfrak{S}(\mathcal{H}')$
be its channel extension.}

\textit{The following properties are equivalent:
}
\begin{itemize}
  \item \textit{there exists an optimal measure for the set
  $\overline{\Phi(\mathfrak{S}(\mathcal{H}))}$;}
  \item \textit{there exists an optimal measure for the channel extension
  $\widehat{\Phi}$.}
\end{itemize}

\textit{The above equivalent properties hold if the set
$\overline{\Phi(\mathfrak{S}(\mathcal{H})}$ contains regular
subset with the same $\chi$-capacity.}

\end{property}

\textbf{Proof.} If $\hat{\mu}_{*}$ is an optimal measure for the
channel extension $\widehat{\Phi}$ then its image
$\hat{\mu}_{*}\circ\widehat{\Phi}^{-1}$ corresponding to the
mapping $\widehat{\Phi}$ is an optimal measure for the set
$\overline{\Phi(\mathfrak{S}(\mathcal{H}))}$.

Let $\nu_{*}$ be an optimal measure for the set
$\overline{\Phi(\mathfrak{S}(\mathcal{H}))}$ and let
$\{\nu_{n}=\{\pi^{n}_{i}, \rho^{n}_{i}\}\}_{n}$ be a sequence of
measures in
$\mathcal{P}(\overline{\Phi(\mathfrak{S}(\mathcal{H}))})$ with
finite support weakly converging to the measure $\nu_{*}$. Since
$\overline{\Phi(\mathfrak{S}(\mathcal{H}))}=\widehat{\Phi}(\widehat{\mathfrak{S}}(\mathcal{H}))$
for each $n$ and $i$ there exists a state $\hat{\rho}^{n}_{i}$ in
$\hat{\mathfrak{S}}(\mathcal{H})$ such that
$\hat{\Phi}(\hat{\rho}^{n}_{i})=\rho^{n}_{i}$. The sequence
$\{\hat{\mu}_{n}=\{\pi^{n}_{i}, \hat{\rho}^{n}_{i}\}\}_{n}$ of
measures in the weakly compact set
$\mathcal{P}(\widehat{\mathfrak{S}}(\mathcal{H}))$ has a weak
limit point $\hat{\mu}_{*}$. Since the mapping $\widehat{\Phi}$ is
continuous with respect to the $*$-weak topology on
$\widehat{\mathfrak{S}}(\mathcal{H})$ and the trace norm topology
on $\mathfrak{S}(\mathcal{H}')$ the image
$\hat{\mu}_{*}\circ\widehat{\Phi}^{-1}$ of the weak limit point
$\hat{\mu}_{*}$ of the set $\{\hat{\mu}_{n}\}_{n}$ is a weak limit
point of the set
$\{\nu_{n}=\hat{\mu}_{n}\circ\widehat{\Phi}^{-1}\}_{n}$, so that
$\hat{\mu}_{*}\circ\widehat{\Phi}^{-1}=\nu_{*}$. Thus
$\hat{\mu}_{*}$ is an optimal measure for the channel extension
$\widehat{\Phi}$.

The last assertion of the proposition follows from the previous
one and theorem 3 in \cite{Sh-4}.$\square$

\begin{remark}\label{on i-o-a-s}
It is clear that each optimal measure for the channel $\Phi$ is an
optimal measure for the channel extension $\hat{\Phi}$.
Proposition \ref{o-m-ext} shows the meaning of the conditions of
existence of an optimal measure for the channel $\Phi$ in theorem
\ref{o-m}. Namely, the regularity condition implies that the set
of optimal measures for the channel extension $\widehat{\Phi}$ is
not empty while the condition of existence of an input optimal
average state for the channel $\Phi$ implies that this set
contains at least one optimal measure for the channel $\Phi$. This
observation is illustrated by the example in the next section.
\end{remark}

\section{On a class of channels with finite $\chi$-capacity}

In this section we consider nontrivial class of
entanglement-breaking channels generalizing the example considered
in \cite{H-Sh-W}. For channels of this class the $\chi$-capacity
and the minimal output entropy can be explicitly calculated. There
also exists simple necessary and sufficient of continuity of the
output entropy for these channels.

Let $G$ be a compact group, $\{V_{g}\}$ be unitary representation
of $G$ on $\mathcal{H}'$, $M(dg)$ be a positive operator-valued
measure (POVM) on $G$, such that the set of probability measures
$\{\mathrm{Tr}\rho M(\cdot)\}_{\rho\in\mathfrak{S}(\mathcal{H})}$
is weakly dense in the set of all probability measures on $G$.

For arbitrary state $\sigma$ in $\mathfrak{S}(\mathcal{H}')$
consider the channel

\begin{equation*}
\Phi_{\sigma}(\rho)=\int_{G}V_{g}\sigma V_{g}^{*}\mathrm{Tr}\rho
M(dg).
\end{equation*}%

Let $\omega(G,V_{g},\sigma)=\int_{G}V_{g}\sigma
V_{g}^{*}\mu_{H}(dg)$, where $\mu_{H}$ is the Haar measure on $G$.
It follows from the assumption of weak density of the set
$\{\mathrm{Tr}\rho M(\cdot)\}_{\rho\in\mathfrak{S}(\mathcal{H})}$
in the set of all Borel probability measures on $G$ that
$\overline{\Phi_{\sigma}(\mathfrak{S}(\mathcal{H}))}=\overline{\mathrm{co}}\{V_{g}\sigma
V_{g}^{*}\}_{g\in G}$. Thus proposition 12 in \cite{Sh-4} implies
the following observation.

\begin{property}\label{orbit-channel}
\textit{The $\chi$-capacity of the channel $\Phi_{\sigma}$ is
equal to
\begin{equation*}
\bar{C}(\Phi_{\sigma})=H(\sigma\|\omega(G,V_{g},\sigma)).
\end{equation*}
If this capacity is finite then
$\Omega(\Phi_{\sigma})=\omega(G,V_{g},\sigma)$ and
$H_{\mathrm{min}}(\Phi_{\sigma})=H(\sigma)$.}

\textit{The channel $\Phi_{\sigma}$ is a $\textup{CE}$-channel if
and only if $H(\omega(G,V_{g},\sigma))<+\infty$. In this case
$\bar{C}(\Phi_{\sigma})=H(\omega(G,V_{g},\sigma))-H(\sigma)
=H(\Omega(\Phi_{\sigma}))-H_{\mathrm{min}}(\Phi_{\sigma})$.}

\end{property}

Since the set
$\overline{\Phi_{\sigma}(\mathfrak{S}(\mathcal{H}))}=\overline{\mathrm{co}}\{V_{g}\sigma
V_{g}^{*}\}_{g\in G}$ is compact for arbitrary $\sigma$
proposition \ref{compact output} implies existence of the
extension $\widehat{\Phi}_{\sigma}$ of the channel $\Phi_{\sigma}$
to the set $\widehat{\mathfrak{S}}(\mathcal{H})$ such that
$
\widehat{\Phi}_{\sigma}(\widehat{\mathfrak{S}}(\mathcal{H}))=\overline{\Phi_{\sigma}(\mathfrak{S}(\mathcal{H}))}
=\overline{\mathrm{co}}\{V_{g}\sigma V_{g}^{*}\}_{g\in G}.
$

Proposition \ref{o-m-ext} and proposition 12 in \cite{Sh-4} show
existence of an optimal measure for the channel extension
$\widehat{\Phi}_{\sigma}$ provided
$\bar{C}(\widehat{\Phi}_{\sigma})=\bar{C}(\Phi_{\sigma})<+\infty$.
The support of any optimal measure for the channel extension
$\widehat{\Phi}_{\sigma}$ consists of states $\hat{\rho}$ in
$\widehat{\mathfrak{S}}(\mathcal{H})$ such that
$\widehat{\Phi}_{\sigma}(\hat{\rho})=V_{g}\sigma V_{g}^{*}$ for
some $g\in G$. As it is shown by the following example we can not
assert existence of an optimal measure for the channel
$\Phi_{\sigma}$ even in the case when $\Phi_{\sigma}$ is a
$\textup{CE}$-channel.

\begin{example}\label{example-2} Consider the case $G=\mathbb{T}$, which can be
identified with $[0,2\pi)$. In this case the Haar measure
$\mu_{H}$ is the normalized Lebesgue measure $\frac{dx}{2\pi}$.
Let $V_{g}$ is the group of shifts in
$\mathcal{H}'=\mathcal{L}_{2}(\mathbb{T})$ and $M(dg)$ is the
spectral measure of the operator of multiplication by an
independent variable in $\mathcal{H}=\mathcal{L}_{2}(\mathbb{T})$.
It is easy to see that the above density assumption is valid in
this case and hence all the above results hold for the
corresponding channel $\Phi_{\sigma}$ with arbitrary
$\sigma\in\mathfrak{S}(\mathcal{L}_{2}(\mathbb{T}))$. Suppose
$\sigma=|\varphi\rangle\langle\varphi|$, where $\varphi$ be an
arbitrary function in $\mathcal{L}_{2}(\mathbb{T})$ with unit
norm. This implies
$\omega(\mathbb{T},V_{g},|\varphi\rangle\langle\varphi|)=(2\pi)^{-1}\int_{0}^{2\pi
}|\varphi_{x}\rangle\langle\varphi_{x}|dx$, where
$\varphi_{x}(t)=\varphi(t-x)$. This channel was originally used in
\cite{H-Sh-W} as an example of entanglement-breaking channel which
has no canonical representation with purely atomic POVM and hence
has no Kraus representation with operators of rank 1. Here we will
use this channel as an example of $\textup{CE}$-channel, which
demonstrates the essential features of infinite-dimensional
channels.

By using proposition \ref{orbit-channel} and direct calculation it
is easy to obtain (cf.\cite{H-Sh-W}) that
\begin{equation}\label{chi-cap-exp}
\bar{C}\left(\Phi_{|\varphi\rangle\langle\varphi|}\right)=
H(\omega(\mathbb{T},V_{g},|\varphi\rangle\langle\varphi|))=
-\sum_{k=-\infty}^{+\infty}|\varphi_{k}|^{2}\log|\varphi_{k}|^{2},
\end{equation}
where $\{\varphi_{k}=(2\pi)^{-1}\int_{0}^{2\pi
}e^{-\mathrm{i}xk}\varphi(x)dx\}_{k\in\mathbb{Z}}$ is the sequence
of the Fourier coefficients of the function $\varphi$, and that
finiteness of $\bar{C}(\Phi_{|\varphi\rangle\langle\varphi|})$
implies $\textup{CE}$-property of the channel
$\Phi_{|\varphi\rangle\langle\varphi|}$ and
$$
\Omega\left(\Phi_{|\varphi\rangle\langle\varphi|}\right)
=\omega(\mathbb{T},V_{g},|\varphi\rangle\langle\varphi|)=
\sum_{k=-\infty}^{+\infty}|\varphi_{k}|^{2}|\tau_{k}\rangle\langle
\tau_{k}|,
$$
where $\{\tau_{k}(x)=\exp(\mathrm{i}kx)\}_{k\in\mathbb{Z}}$ is the
trigonometric basis in $\mathcal{L}_{2}(\mathbb{T})$. It is
interesting to note that the properties of the channel
$\Phi_{|\varphi\rangle\langle\varphi|}$ are determined by the rate
of vanishing of the Fourier coefficients of the function
$\varphi$.

Suppose series (\ref{chi-cap-exp}) is finite for a function
$\varphi$ and hence $\Phi_{|\varphi\rangle\langle\varphi|}$ is a
$\textup{CE}$-channel. By proposition \ref{orbit-channel}
$H_{\mathrm{min}}(\Phi_{|\varphi\rangle\langle\varphi|})=0$ in
spite of the fact that the set
$\Phi_{|\varphi\rangle\langle\varphi|}(\mathfrak{S}(\mathcal{H}))$
does not contain pure states. The set
$\{|\varphi_{x}\rangle\langle\varphi_{x}|\}_{x\in\mathbb{T}}$ of
pure states in
$\overline{\Phi_{|\varphi\rangle\langle\varphi|}(\mathfrak{S}(\mathcal{H}))}$
is contained in the output set of the channel extension
$\widehat{\Phi}_{|\varphi\rangle\langle\varphi|}$ and corresponds
to a particular subset of nonnormal states in
$\widehat{\mathfrak{S}}(\mathcal{H})$. By proposition
\ref{chi-function} the functions
$\chi_{\Phi_{|\varphi\rangle\langle\varphi|}}$ and
$\hat{H}_{\Phi_{|\varphi\rangle\langle\varphi|}}$ are bounded and
continuous on $\mathfrak{S}(\mathcal{H})$. Nevertheless there
exists no optimal measure and hence there exists no input optimal
average state for the channel
$\Phi_{|\varphi\rangle\langle\varphi|}$ (since otherwise corollary
\ref{o-m-s-c} implies existence of an optimal measure). The
continuous functions
$\chi_{\Phi_{|\varphi\rangle\langle\varphi|}}$ and
$\hat{H}_{\Phi_{|\varphi\rangle\langle\varphi|}}$ do not achieve
their finite supremum
$\bar{C}(\Phi_{|\varphi\rangle\langle\varphi|})$ and infinum
$H_{\mathrm{min}}(\Phi_{|\varphi\rangle\langle\varphi|})=0$
correspondingly on noncompact set $\mathfrak{S}(\mathcal{H})$.

The family of channels
$\{\Phi_{|\varphi\rangle\langle\varphi|}\}_{\varphi\in\mathcal{L}_{2}(\mathbb{T})}$
provides an another example showing that in the infinite
dimensional case the $\chi$-capacity is not continuous but only
lower semicontinuous function of a channel \cite{Sh-2}.

It is easy to see that for arbitrary unit vector
$|\varphi_{*}\rangle$ in $\mathcal{L}_{2}(\mathbb{T})$ and for
arbitrary sequence $\{|\varphi_{n}\rangle\}$ of unit vectors in
$\mathcal{L}_{2}(\mathbb{T})$ converging to the vector
$|\varphi_{*}\rangle$ the sequence
$\sup_{\rho\in\mathfrak{S}(\mathcal{H})}\|\Phi_{|\varphi_{n}\rangle\langle\varphi_{n}|}(\rho)-\Phi_{|\varphi_{*}\rangle\langle\varphi_{*}|}(\rho)\|_{1}$
tends to zero. It means that
$\varphi\mapsto\Phi_{|\varphi\rangle\langle\varphi|}$ is a
continuous mapping from $\mathcal{L}_{2}(\mathbb{T})$ into the set
of all channels endowed with the topology of uniform convergence.

Let $\{\varphi_{n}\}$ be the sequence of function in
$\mathcal{L}_{2}(\mathbb{T})$ with the following Fourier
coefficients
$$
   (\varphi_{n})_{k}=\left\{
   \begin{array}{ll}
    \sqrt{1-q_{n}},& k=0\\
    \sqrt{q_{n}/n}, & 0<k\leq n \\
    \;0, & \;\textup{for}\;\textup{others}\;k,
    \end{array}\right.
$$
where $\{q_{n}\}$ is a sequence of numbers in $(0,1)$ such that
$\lim_{n\rightarrow+\infty}q_{n}\log n=C>0$. It is easy to see
that the sequence $\{\varphi_{n}\}$ converges in
$\mathcal{L}_{2}(\mathbb{T})$ to the function
$\varphi_{*}(x)\equiv 1$. The above observation implies that the
sequence $\{\Phi_{|\varphi_{n}\rangle\langle\varphi_{n}|}\}$ of
channels uniformly converges to the channel
$\Phi_{|\varphi_{*}\rangle\langle\varphi_{*}|}$, for which
$\Phi_{|\varphi_{*}\rangle\langle\varphi_{*}|}(\rho)=|\varphi_{*}\rangle\langle\varphi_{*}|$
for all $\rho$ in $\mathfrak{S}(\mathcal{H})$ and hence
$\bar{C}(\Phi_{|\varphi_{*}\rangle\langle\varphi_{*}|})=0$. But by
(\ref{chi-cap-exp}) we have
$$
\bar{C}\left(\Phi_{|\varphi_{n}\rangle\langle\varphi_{n}|}\right)=-q_{n}\log
q_{n}-(1-q_{n})\log(1-q_{n})+q_{n}\log n,\quad \forall n\in
\mathbb{N},
$$
and hence
$\lim_{n\rightarrow+\infty}\bar{C}\left(\Phi_{|\varphi_{n}\rangle\langle\varphi_{n}|}\right)=C>0$.

\end{example}

\textbf{Acknowledgments.} The author is grateful to A. S. Holevo
for the help in preparing of this paper. The work was partially
supported by the program "Modern problems of theoretical
mathematics" of Russian Academy of Sciences.

\end{document}